\documentclass[prb,aps,10pt,amsmath,amssymb,floatfix,twocolumn,superscriptaddress]{revtex4-1}

\usepackage[english]{babel}
\usepackage[utf8x]{inputenc}
\usepackage{graphicx}
\usepackage{amsmath}
\usepackage{amssymb}
\usepackage{hyperref}
\usepackage{xspace}
\usepackage{gensymb}

\let\AAt\AA
\renewcommand{\AA}{\ifmmode{\mathrm{\AAt}}\else{\AAt}\fi\xspace}

\newcommand{\unit}[1]{\ensuremath{\mathrm{#1}}\xspace}

\newcommand{\chem}[1]{\ensuremath{\rm #1}\xspace}
\newcommand{\Ti}{\chem{Ti}}
\newcommand{\mB}{\ensuremath{\rm \mu_B}\xspace}
\newcommand{\eV}{\ensuremath{\rm eV}\xspace}
\newcommand{\meV}{\ensuremath{\rm meV}\xspace}
\newcommand{\hole}{\ensuremath{\mathrm{\mathbf{h}}}\xspace}
\newcommand{\el}{\ensuremath{\mathrm{\overline{e}}}\xspace}

\newcommand{\STO}{\chem{SrTiO_3}}

\newcommand{\OV}{\chem{V_{\!O}}}

\preprint{SrTiO3 Vacancy Magnetism Control v2.5 - 25/08/2017}

\begin{document}

	\title{Controlling the magnetism of oxygen surface vacancies in \STO through charging}
	
	\author{Oleg~O.~Brovko}
	\email{obrovko@ictp.it}
	\affiliation{The Abdus Salam International Centre for Theoretical Physics (ICTP), Trieste (TS), Italy}
	\author{Erio~Tosatti}
	\email{tosatti@sissa.it}
	\affiliation{The Abdus Salam International Centre for Theoretical Physics (ICTP), Trieste (TS), Italy}
	\affiliation{Scuola Internazionale Superiore di Studi Avanzati, Trieste (TS), Italy}
	
	\begin{abstract}
		We discuss, based on first principles calculations, the possibility to tune the magnetism of oxygen vacancies at the (001) surface of strontium titanate (\STO). The magnetic moment of single and clustered vacancies stemming from $\chem{Ti-O}$ broken bonds can be both quenched and stabilized controllably by chemical potential adjustment associated with doping the system with electrons or holes. We discuss to what extent this route to magnetization state control is robust against other external influences like chemical doping, mechanical action and electric field. Such control of vacancy state and magnetization can conceivably be achieved experimentally by using local probe tips. 
	\end{abstract}

	\maketitle

\section*{Introduction}

	Throughout decades strontium titanate (\STO) has continuously been in the spotlight of solid state research as a material with rich and varied physics. More recently it also became the substrate of choice for research and applications in the field of oxide electronics.~\cite{McKee1998,Goodenough2004,Marshall2015} \STO interfaces were shown to support a 2D electron gas with extremely high electron mobility values and diverse superconducting and relativistic physics.~\cite{Ohtomo2004,Ohta2007,Santander-Syro2014,Plumb2014,Taniuchi2016} Strain-controlled room temperature ferroelectricity,~\cite{Haeni2004} a useful trait for device applications, has been shown to exist in \STO alongside quantum paraelectricity at low temperatures.~\cite{Muller1979, Muller1991} With the advent of spintronics \STO has emerged as a wide band gap insulator material capable of itinerant, impurity and vacancy based magnetism.~\cite{Marshall2015} It has been long known, that bulk impurities~\cite{Blazey1983,Bannikov2008,Liu2015a} in \STO can be used to tailor the oxide's optical and electronic properties. Even in the absence of foreign atoms, pristine \STO is prone to forming oxygen vacancies (\OV) if annealed at higher temperatures under oxygen-poor conditions,~\cite{Klie2000,Janousch2007} when bombarded with noble gas ions~\cite{Chang2015} or under intense laser or ultraviolet irradiation.~\cite{Rao2014,Zhang2015} Oxygen vacancies in \STO bulk were studied extensively both theoretically~\cite{Selme1983,Shanthi1998,Ricci2003,Shein2007a,Kotomin2008,Kim2009,Hou2010,Liao2012,Lin2012b,Lin2013a,Pavlenko2012,Gryaznov2013,Lopez-Bezanilla2015,Lopez-Bezanilla2015a,Zhang2015a} and experimentally.~\cite{Klie2000,Muller2004a,Kalabukhov2007,Janousch2007,Kim2009,Jiang2011,Middey2012,DeSouza2012,Rao2014,Rice2014,Rice2014a,Chang2015,Zhang2015,Trabelsi2016} They were shown not only to represent the key to metalization and control over the carrier density and mobility in \STO, but also to be inherently magnetic.~\cite{Shanthi1998,Ricci2003,Shein2007a,Hou2010,Liao2012,Middey2012,Lin2013a,Rao2014,Rice2014,Rice2014a,Lopez-Bezanilla2015,Lopez-Bezanilla2015a,Zhang2015a,Trabelsi2016} Importantly for spintronic applications and nanoscale surface studies, similar trends for oxygen vacancies were found at \STO surfaces.~\cite{Crandles2010,Pavlenko2013,Xu2013,Li2015,Garcia-Castro2016,Taniuchi2016,Altmeyer2016,Kimura1995,Stashans2002,Cai2006,Alexandrov2009,Alexandrov2009a,Zhukovskii2009,Shen2012a,Choi2013} Depending on the concentration~\cite{Pavlenko2013,Lopez-Bezanilla2015} and clustering patterns~\cite{Cuong2007,Hou2010,Liao2012,Lopez-Bezanilla2015a,Li2015} \OV were shown to exhibit either local uncorrelated~\cite{Lin2013a,Garcia-Castro2016} magnetic signatures or, when sufficiently abundant, a long-range and stable magnetic order.~\cite{Liao2012,Pavlenko2013,Rice2014,Rice2014a,Taniuchi2016,Altmeyer2016,Trabelsi2016} As abundant as the existing pool of literature on surface \OV in \STO is, it is also equally controversial, especially its theoretical component. Partially it is due to the limitations of the supercell computational approach, mainly used for crystal surface calculations. As our study has shown relatively large cells (typically at least 3-4 units of \STO laterally) are required to quench the spurious interaction between the impurities/vacancies. In view of this we do not attempt to survey all the existing claims and assess their veracity referring the reader instead to the abundant above list of citations and concentrating instead on a different issue, namely the possibility to tailor the magnetic properties of the \OV (the magnetic nature thereof is almost unanimously accepted).  

	Some degree of control over \OV-related magnetism was shown to be achievable through external or interface stress~\cite{Zhang2015a}, but the quest for an effective magnetization tuning mechanism is still on. Non-magnetic investigations of the \OV have repeatedly demonstrated that the charge state of the vacancy has, as it is natural, a pronounced effect on its electronic and structural properties. The aim of the present work is to address the related important questions, namely: (i) how does charging of oxygen vacancies at the (100) surfaces of \STO affect in detail its magnetic properties?; (ii) under what conditions would this charging and control be achievable?; and lastly (iii) how robust can this effect be against external influences, such as mechanical action, doping or electric field exposure? 

	We present calculations and arguments showing that oxygen vacancies at, and close to, the (100) surface of \STO can be indeed inherently magnetic depending on the charging state, and in addition the extent to which this magnetism is robust against structural changes in the atomic arrangement, doping and electric fields. Model results also highlight the way in which the magnetic state of vacancies and vacancy clusters can be either turned or quenched by externally induced charging.


\section*{Methods and Geometries}
	
	First principles calculations were carried out in the framework of the density functional theory (DFT), based on the projector-augmented-wave method,~\cite{Blochl1994} and a plane-wave basis set~\cite{Kresse1996} as implemented in the Vienna Ab-initio Simulation Package (VASP).~\cite{Kresse1993,Kresse1996} Exchange and correlation were treated with the gradient-corrected functional as formalized by Perdew, Burke and Ernzerhof.~\cite{Perdew1996} On-site Coulomb interaction corrections were accounted for in the framework of the LSDA+U formalism as introduced by Dudarev \textit{et al.}~\cite{Dudarev1998} The values for the Hubbard $U$ and $J$ parameters for the Ti $d$-orbitals were taken to be $5~\unit{eV}$ and $0.64~\unit{eV}$ respectively after Ref.~\onlinecite{Okamoto2006} and checked against instability with $U=4~\unit{eV}$ after Ref.~\onlinecite{Mizokawa1995}. For bulk and lattice constant calculations an energy cutoff of $600~\unit{eV}$ for the plane wave expansion and a Monkhorst-Pack $k$-point mesh~\cite{Monkhorst1976} with $29\!\times\!29\!\times\!29$ points (before symmetry operations application) were used. The repeated-cell geometry for bulk vacancy calculations was taken to comprise $4\!\times\!3\!\times\!3$ \STO unit cells. For surface calculations a slab of 4 \STO unit layers was used where two units were fixed at bulk geometry and the top two were allowed to fully relax. In the $z$ direction the slabs were separated from their periodic images by $15\AA$ of vacuum. For vacancy calculation atoms within two atomic shells from the vacancy site were allowed to relax. Obtained relaxations are in line with those found in similar studies,~\cite{Zhukovskii2009,Alexandrov2009,Alexandrov2009a,Janotti2014} namely generally outward vertical and inward (towards the impurity) relaxation of surface oxygen atoms surrounding the vacancy and outward in-plane relaxation of Ti atoms (depending on the charge state). The relaxations of Ti atoms around a neutral vacancy is a controversial issue and the reported relaxations vary among the above publications. We find an outward relaxation to be the energetic ground state. Moreover we find that the answer is sensitive to the size of the calculation cell, wich in our case was larger than in any of the previous studies. 

	For each vacancy state,  relevant quantities are calculated including among others, the formation energy of a $\nu$-atom oxygen vacancy with $q$ electrons ($q=0$ being the neutral vacancy)
	\begin{equation}
		E_{\mathrm form}(\nu, q) = E_{\nu}^q  + \nu \cdot E_{\chem{O_2}}/2 - E_{0}^q,\label{eq:eform}
	\end{equation}
	as well al the work function
	\begin{equation}
		\Phi = \epsilon_0 - E_{\mathrm F}
	\end{equation}
	where $E_{\mathrm F}$ is the one-electron Fermi level position within Kohn-Sham eigenvalues, $\epsilon_0$ is the vacuum zero  extrapolated as far above the surface as possible, and $E_{\nu}^q$, $E_0^q$ and $E_{\chem{O_2}}$ are the total energies of a $\nu$-atom oxygen vacancy, a clean surface (with $q$ electrons) and an diatomic oxygen molecule in gas phase respectively. The stable charge state of a vacancy or vacancy cluster will correspond to the lowest value of the grand potential $E_{\mathrm form}(\nu, q) - \mu \cdot q$. Here $\mu$ is the chemical potential, whose difference from vacuum zero coincides with the work function $\Phi$ in the absence of external fields, becoming in our context a free variable controlled  doping, or by the external potential of a tip, etc.

	For the calculation of a reconstructed surface and surface-based vacancies the supercell of the calculation consisted of $3\!\times\!3$ and $4\!\times\!5$ \STO unit cells in plane of the surface. The $k$-point mesh used in this case was $3\!\times\!3\!\times\!1$ and $\Gamma$-point-only respectively.
	
	Most of the conclusions derived in the present paper are based on calculations carried out for vacancies residing on, or close to, non-reconstructed \chem{TiO} and \chem{SrO} terminated \STO surfaces. We deliberately chose to neglect the tetragonal antiferrodistortive phase of \STO below the transition temperature of $\sim 105\!-\!110\degree\mathrm{C}$,~\cite{Muller1969} since the latter is known to have a limited effect on the electronic properties of \STO~\cite{Choi2013} while neglecting it allows for a higher degree of generality and transferability in first principle calculations. From numerous experimental and theoretical studies it is known that the \chem{TiO} termination is prevalent and energetically slightly more stable under ambient conditions,~\cite{Padilla1998,Piskunov2005} though SrO terminated surfaces can be easily produced by growth in \chem{Sr}-rich atmosphere or controlled hydroxylation.~\cite{Koster1998} Moreover, pristine surfaces of \STO often reconstruct exhibiting a broad range of geometries, $2\!\times\!1$, $2\!\times\!2$, $c(4\!\times\!3)$, $c(6\!\times\!2)$, $\sqrt{5}\!\times\!\sqrt{5}-R26.6\degree$ and $\sqrt{13}\!\times\!\sqrt{13}-R37.7\degree$ being the the most common ones.~\cite{Newell2007} To test the validity of our results for realistic reconstructed surfaces we chose, guided by discussions with M. Kisiel \footnote{private communication}, to investigate oxygen vacancies at several proposed realizations of one particular reconstruction pattern, namely a $2\!\times\!2$ one.~\cite{Shiraki2010,Lin2011a}

\section*{Results and Discussion}


\subsection*{Single surface vacancy}
	
	\begin{figure}
		\center{\includegraphics{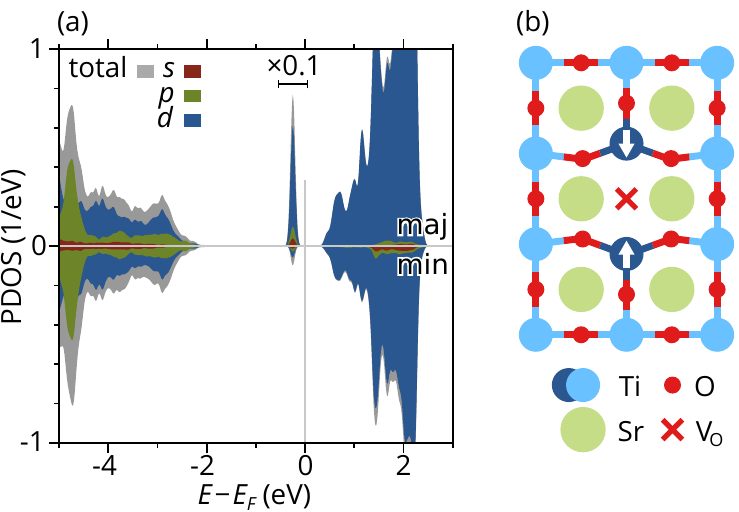}}
		\caption{Partial density of states (a) of one of the two equivalent \Ti atoms [marked deep blue in the sketch of the system in panel (b)] neighboring a single surface oxygen vacancy. Red, green and blue filled curves show the $s$, $p$ and $d$ \Ti-orbital projected contributions respectively. Positive: majority (spin up), negative: minority (spin down). The spin is almost entirely localized in the deep-gap bound state just below $E_{\rm F}$. The magnetization sign of the two \Ti atoms is opposite, so that the total magnetization of the vacancy is zero by symmetry. Panel (b) contains the sketch of the system with red circles denoting oxygen and blue ones standing for titanium. Deep blue marks the Ti atoms shouldering the vacancy. Arrows therein denote the spin orientation in the ground state magnetic configuration.}
		\label{fig:dos:v1}
	\end{figure}
	
	We start our investigation with examining a single oxygen vacancy at a \chem{TiO} surface of $\STO(001)$.~\footnote{A brief summary of a benchmark calculation of the bulk vacancy can be found in the Supplemental Material Fig.~S1} Our calculation  yields, in accord with the extensive existing literature pool, a ground state with excess charge of the vacancy localized at the \Ti atoms neighboring the \OV site [see the geometry sketch in Fig.~\ref{fig:dos:v1}(b)]. The $d$-orbitals accommodating the charge form an impurity level deep inside the electronic band gap of the \STO surface as is illustrated by the partial density of states [PDOS, Fig.~\ref{fig:dos:v1}(a)] of one of the \Ti atoms neighboring the vacancy [shaded dark blue in Fig.~\ref{fig:dos:v1}(b)]. The two excess electrons left behind by the departed \chem{O} atom cause the \Ti  atoms to acquire magnetic moments of $1\mB$ each, localized in their $d_z$ \Ti-orbital. The two \Ti spins facing each other across the vacancy are antiferromagnetically coupled with an exchange energy of about $750~\meV$, (as measured by the energy difference  $E_{\rm tot}(m=2\mB) - E_{\rm tot}(m=0\mB)$) so that the total vacancy magnetization is zero.
	
	\begin{figure}
		\center{\includegraphics{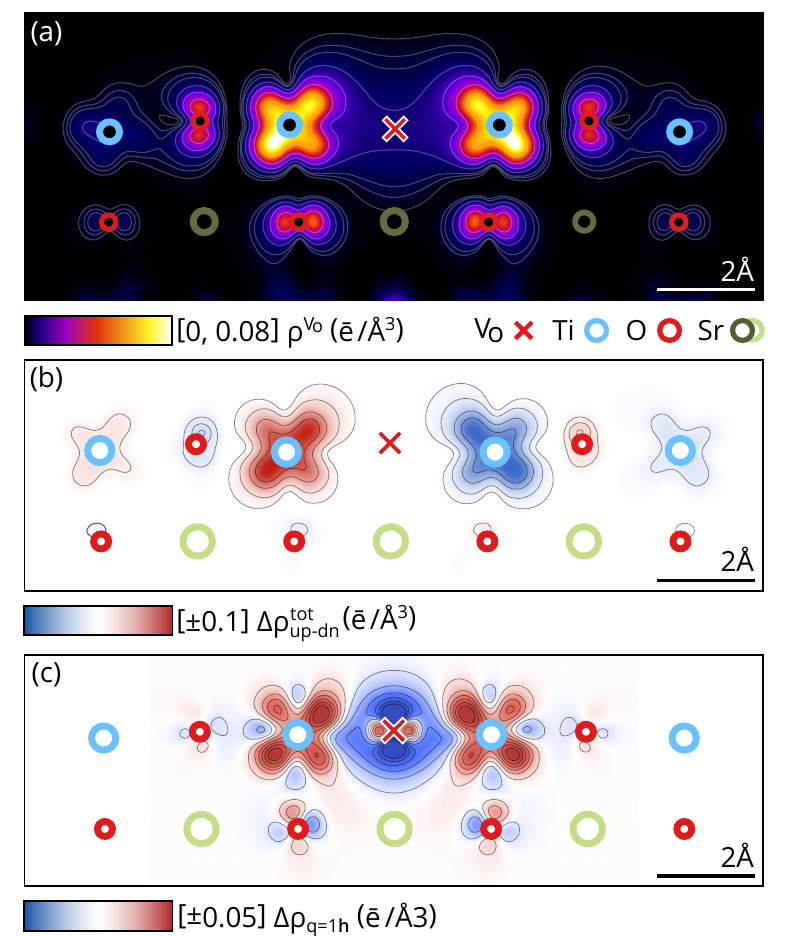}}
		\caption{(a) Spatial charge density of the Kohn-Sham oxygen vacancy level in the gap shown in Fig.~\ref{fig:dos:v1}(a) -- cross-section by a plane normal to the \STO surface passing through the vacancy site and the two neighboring \Ti atoms. This charge distribution, closely corresponding to the modulus of the spin density distribution, is predominantly localized at the \Ti atoms, with a partial spill-over onto the neighboring oxygen sites. (b) Corresponding electron spin asymmetry (note that the that the color scale here is non-linear, the actual spin asymmetry on the sites neigboring the Ti atoms is almost negligibly small). (c) Charge redistribution pattern, calculated as $\Delta\rho = \rho_{\OV} + \rho_{\chem{O}} - \rho_{\chem{clean}}$, where $\rho_{\OV}$, $\rho_{\chem{clean}}$ and $\rho_{\chem{O}}$ are the charge density distributions of the surface slab with and without a vacancy and a free-standing oxygen atom respectively.}
		\label{fig:chredist}
	\end{figure}
	
	To visualize the strength of localization of the charge trapped by the vacancy at the neighboring \Ti atoms we examine the spatial charge density of the Kohn-Sham states corresponding to the \OV level. Fig.~\ref{fig:chredist}(a) shows a cut of the above charge density by a plane normal to the surface and passing through the vacancy site and the neighboring \Ti atoms. It is apparent that the majority of the electron density of this state is concentrated in the Ti-$d$ orbitals with some of it spilling over to the neighboring oxygens. The antiferromagnetic alignment of \Ti spins can be clearly observed if we visualize the electron spin density by plotting the spin asymmety $P=\rho^{\uparrow}-\rho^{\downarrow}$ ($\rho^{\uparrow,\downarrow}$ are the densities of majority and minority electrons respectively) of electrons in the same plane as shown in Fig.~\ref{fig:chredist}(a). The antisymmetry of the map is a usual signature of antiferromagnetism. The atoms in the first neigbor shell of the magnetic \Ti atoms acquire a small induced magnetic moment, but note that the color scale here is non-linear, so that the absolute spin polarization value on those sites is almost negligibly small.
	
	Another way of visualizing the charge contents of the oxygen vacancy is looking at the charge density redistribution caused by its creation. In Fig.~\ref{fig:chredist}(c) we plot the charge redistribution $\Delta\rho = \rho_{\OV} + \rho_{\chem{O}} - \rho_{\chem{clean}}$, where $\rho_{\OV}$, $\rho_{\chem{clean}}$ and $\rho_{\chem{O}}$ are the charge density distributions of the surface slab with and without a vacancy and a free-standing oxygen atom respectively [the cross-section plane is the same as for Fig.~\ref{fig:chredist}(a)]. It represents visually the transfer of the electrons formerly attached to oxygen (nominally two) from the site now vacant to the $d$-orbitals of the neighboring \Ti atoms.
	
	\begin{figure}
		\center{\includegraphics{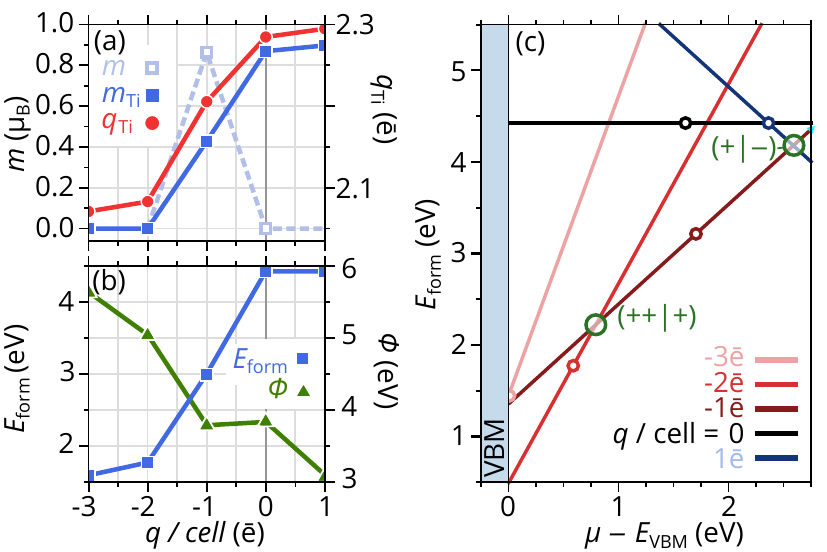}}
		\caption{Electron and hole doping dependence of the (a) Bader charge $q_{\Ti}$ (red circles) and magnetic moments' magnitude $m_{\Ti}$ (blue squares) of each of the \Ti atoms neighboring the oxygen vacancy and the total vacancy moment $m$, (b) the formation energy $E_{\rm form}$ of the vacancy (blue squares) and the work function (chemical potential $\mu$ at $q$=0 ) of the electrons at the Fermi level (in the vacancy level) for a single oxygen vacancy at the $\STO(001)$ surface. (c) Formation energy diagram for different charge states of a single oxygen vacancy at the \chem{TiO}-terminated $\STO(001)$ surface as a function of the chemical potential [given with respect to the valence band maximum (VBM)]. The value $m_{\Ti} = 1/2\mB$ for $q =-1\el$ reflects a single electron spin shared between the two \Ti atoms In this case, the total electron number being odd, the total vacancy magnetization is $m=1\mB$.}
		\label{fig:v1}
	\end{figure}
	
	Since the magnetization of the \Ti atoms is directly linked to the excess charge conveyed to them by creation of the vacancy electrons it stands to reason that the magnetization state should be highly susceptible to charge doping and/or depletion in the system. To verify that we calculate the ground state of the system (the largest supercell considered here, \textit{i.e.} the $4\!\times\!5$ slab) with the charge $q$ of the cell increased or reduced by an integer number of electrons $n_{\el}$. On account of the nonmagnetic and insulating character of bulk \STO the total magnetization $m$ is set to $1\mB$ for odd $n_{\el}$, and $m= 0$ otherwise. Higher magnetizations were also explored and found to be energetically unfavorable with respect to the lower ones. The vacancy electronic structure of Fig.~\ref{fig:v1}(a) makes it clear why. There are only two midgap \Ti-broken bond levels in the insulating gap; magnetizations larger than $1~\mB$ involve additional promotion of electrons/holes from midgap to the conduction or valence-derived bands and are thus energetically more costly.~\footnote{Note, however, that prevalence of low magnetizations is not universal and does not generally hold for extended or interacting oxygen vacancies in \STO.} Fig.~\ref{fig:v1}(a) shows the evolution of the magnetic moments' magnitude $|m_{\Ti}|$  (blue squares) and the valence\footnote{The ``core'' electrons are excluded from the summation.} Bader charge~\cite{Bader1985,Tang2009}  $q_{\Ti}$ (red circles) of each of the two \Ti atoms neighboring the \OV as we vary the number of electrons. While lowering of the chemical potential (increasing number of electrons) barely has an effect on the already occupied impurity level localized at \Ti atoms (their Bader charge remains unchanged as the cell is negatively charged), an increase in chemical potential causes the impurity level to gradually deplete, resulting concurrently in a reduction of  $m_{\Ti}$.
	
	The corresponding change of the work function in \eV and the vacancy formation energy change are plotted in Fig.~\ref{fig:v1}(b) (green triangles and blue squares respectively). Note that in the numerical approach used in the present study the chemical potential is altered by constraining the number of electrons in the calculation cell, the actual change of the Bader charge of the near-vacancy \Ti atoms associated with integer electron addition or depletion is relatively small (fractions of an electron), which is nonetheless sufficient to completely quench their magnetic moment. The remaining charge is detracted from the surrounding atoms with non-zero occupation of the vacancy level (visually the extent of impurity level can be estimated from Fig.~\ref{fig:chredist}(a)). This demonstrates that control over magnetism in oxygen vacancies at \STO surfaces can be achieved through chemical potential tailoring (charge injection/depletion). The magnetic coupling between the near-vacancy \Ti is found to be antiferromagnetic as shown by the total moment $m$ of the impurity alternating between zero for an uncharged vacancy and $1~\mB$ for $|n_{\el}|=1\el$. 
	
	The next important thing to consider, however, is that while our constrained density functional calculation does yield a ground state solution for each charge state of the finite periodic system, it does not automatically imply that the state shall be the ground state of an isolated impurity in a real-life \STO sample, where the result would be determined by the position of the chemical potential. To examine the landscape of achievable charge states we plot in Fig.~\ref{fig:v1}(c) the formation energy (Eq.~\ref{eq:eform})~\footnote{The formation energy of the vacancy in a given charge state is calculated as the total energy difference between the cell with an oxygen vacancy and a sum of clean \STO surface energy and the chemical potential of an oxygen atom in a free-standing oxygen molecule.} diagram for different charge states of a single oxygen vacancy at the \chem{TiO}-terminated $\STO(001)$ unreconstructed surface as a function of the chemical potential (given with respect to the valence band maximum (VBM). The markers on the lines represent the constrained-charge ground state given by DFT. The green circles represent the transition points and are annotated to mark the charge states between which the system switches (with ``$+$'' and ``$-$'' denoting the number of holes in the system with respect to the neutral state ``0''). At any given chemical potential the lowest line in the diagram defines the preferred charge state of the oxygen vacancy. Somewhat unexpectedly, our calculations indicate that the neutral state of the vacancy is not achieved at any chemical potential. In its place, a direct transition from the singly positively charged to singly negatively charged state is preferred. In the real world this would amount to one electron of the vacancy in a neutral state being donated to the host matrix. While this is in line with numerous other theoretical predictions of oxygen vacancies being responsible for the formation of an itinerant electron gas at the surface of \STO, it has also to be considered, that the range of chemical potentials where a neutral vacancy is closest to becoming the ground state is very close to the width of the band gap,~\cite{Zhukovskii2009} which is underestimated by the DFT by almost an \eV. Correcting the size of the gap in the calculation might yield a slightly different formation energy diagram from that depicted in Fig.~\ref{fig:v1}(c), perhaps with a range of chemical potentials allowing for an uncharged ground state of the \OV. However, the above mentioned gap correction in DFT would require employing hybrid exchange and correlation functionals which unfortunately is computationally too heavy for the system sizes considered here. Nonetheless, the qualitative statement that a magnetic state of an oxygen vacancy (as a quantum dot) is susceptible to manipulation through chemical potential engineering still stands.
	
	To underscore and confirm the generality of the above conclusion we examine first of all several other configurations of single oxygen vacancies at \STO surfaces. We calculate single vacancies in the second and third layers of the \chem{TiO_2}-terminated surface, as well as \OV in the first and second layers of the \chem{SrO}-terminated \STO crystal. All cases exhibit similar traits, \textit{i.e.} magnetic \Ti ion pairs at the vacancy site in its neutral state and responsiveness to charge manipulation leading to magnetic moment reduction and quenching following the drop of chemical potential.
		
	A valid question to address at this point would be the scope of mechanisms available to achieve the chemical potential tuning, \textit{i.e.} external agents capable of changing the local chemistry in the vicinity of the oxygen vacancy. Addressing all of them would spring the limits of the present study, but we would like to specifically mention one straightforward way of injecting charge (both electrons and holes) into a surface vacancy which is particularly relevant in the framework of contemporary spintronic applications, namely to approach the \OV with a local probe (such as an STM or AFM tip) and possibly create a potential gradient (electric field) giving rise the transfer/tunnelling of electrons between the tip and the vacancy. While it is also conceivable~\cite{Li2013c, Brovko2014, Shimizu2015, Coey2016} to use the effect of the electric field, doping or mechanical action to tune the charge and magnetic properties of surface defects, our test calculations involving exposure of a \OV at a \Ti-terminated \STO surface to electric fields up to $1~\eV/\AA$ and mechanical lift-off of the topmost layer with forces in excess of those achievable by Van der Waals forces in junction geometry have shown that neither of the latter has a significant effect on the charge state or indeed the magnetic properties of an oxygen vacancy. 

	This means, on one hand, that the choice of tools for harnessing oxygen vacancies as magnetic quantum dots it limited, but on the other hand it indicates that directly addressing \OV-s with a local probe tip is a viable channel of spin manipulation not susceptible to environmental instabilities. Note, however, that in the present study the change in electron number, as attainable in real life by the external potential exerted by a tip, has been forced onto a system devoid of free charge carriers so that no other effects of the external potential are expected apart from the local chemical potential change. In real \STO samples the accumulation of impurities at the surface can cause a non-zero population of free or semi-free surface electrons and so can the surface state.~\cite{Santander-Syro2011,Meevasana2011,Shen2012a,Marshall2015,Coey2016} Those electrons can be redistributed by the application of a bias via a local probe tip leading to a population or depletion of a vacancy site under the tip and thus a change of its magnetic moment. This scenario can actually be seen as the main practical instrument to alter the chemical potential locally leading to the alteration of the impurities' spin state as described above.

	
\subsection*{Vacancy clusters}

	\begin{figure}
		\center{\includegraphics{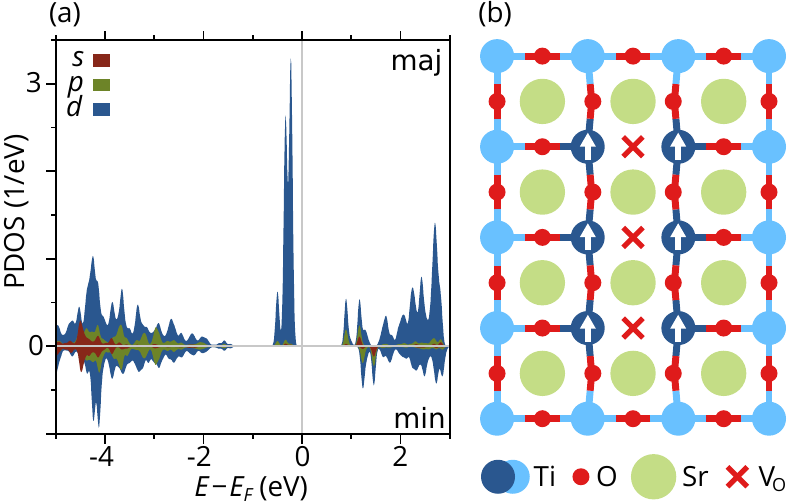}}
		\caption{(a) Partial density of states of one of the central \Ti atoms neighboring a triple oxygen vacancy as shown in panel (b). Red, green and blue filled curves show the $s$, $p$ and $d$ orbital projected contributions respectively. Positive: majority (spin up), negative: minority (spin down). Panel (b) contains the sketch of the system with red circles denoting oxygen and blue ones standing for titanium. Deep blue marks the Ti atoms shouldering the vacancy. Arrows therein denote the spin orientation in the ground state magnetic configuration.}
		\label{fig:dos:v3b}
	\end{figure}
	
	So far we described charge and magnetization switching in a single, isolated oxygen vacancy. Both experimental and theoretical evidence however point towards vacancy migration to surfaces and interfaces and their aggregation into clusters.~\cite{Alexandrov2009,Marshall2015,Li2015} We therefore proceed to carry out  calculations for oxygen vacancy clusters. In view of controversial claims in the literature concerning the nature of in-plane vacancy clustering, see e.g.~\onlinecite{Jeschke2015}, we seek to further increase the generality of our conclusions by studying several representative two-dimensional vacancy clusters residing in the topmost surface layer (see Supplemental Material Fig.~S2 for the list of cluster configurations studied) for the signatures of magnetism and charge state transitions. We find that all the vacancy clusters studied exhibit ground state magnetization of \Ti atoms neighboring the \OV sites. Moreover, for most non-linear vacancy cluster configurations the coupling between the \Ti atom spins is antiferromagnetic, yielding either $m=0$ ($m=1\mB$) net magnetic moment of the vacancy cluster as a whole for even (odd) electron number $n_{\el}$. Similarly to the case of a single vacancy, electron depletion causes the vacancy cluster to lose the charge localized therein and the moment of the \Ti atoms to be gradually quenched. 
We consider however that previous studies, also confirmed by our calculations, predict the formation of stable linear vacancy chains~\cite{Cuong2007} as an important mode of vacancy clustering. We address this possibility by studying as an illustrative example the case of a linear vacancy chain consisting of 3 contiguous oxygen vacancies. Our calculations show that similar to other linear vacancy arrays spins on near-vacancy \Ti atoms are in this case ferromagnetically coupled therefore stabilizing the maximum $m$ value at any given charge state $n_{\el}$. The neutral state of the triple linear vacancy has therefore  $m= 6\mB$. The sketch of the system is presented in Fig.~\ref{fig:dos:v3b}(b) and the partial projected DOS of one of the \Ti atoms neighboring the middle oxygen vacancy is shown in panel (a) of the same figure. The presence can be noted of several \Ti-connected levels in the vacancy cluster gap. The apparent broadening caused by their crowding in the, as well as by the unphysical overlap due to periodic boundary conditions, preserves the magnetic character thereof in some cases raising the vacancy magnetization from lowest to highest.
	
	\begin{figure}
		\center{\includegraphics{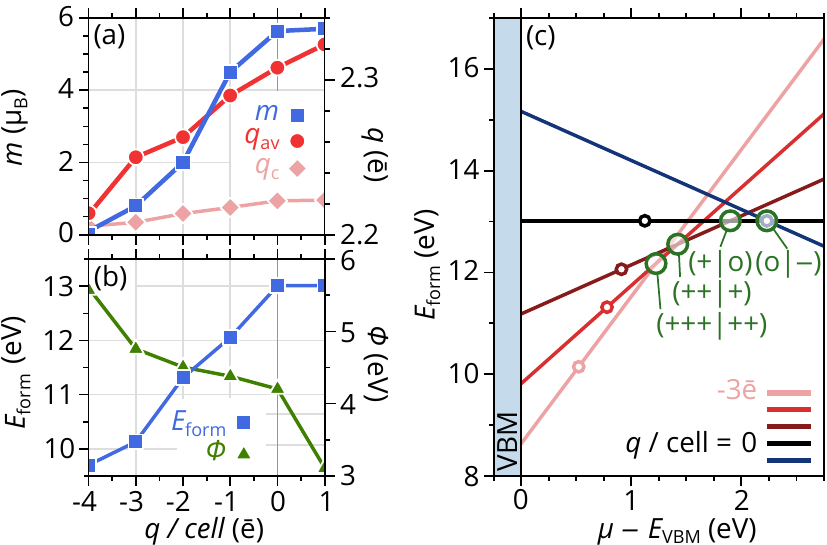}}
		\caption{Electron and hole doping dependence of the (a) valence Bader charge (red circles for average charge $q_{\rm av}$ of the \Ti atoms neighboring the chain [shaded dark blue in panel (b)] and pale red rhombs for the charge $q_{\rm c}$ on one of the \Ti atoms neighboring the middle oxygen vacancy) and cumulative magnetic moment $m$ (blue squares) of the \OV chain, (b) the formation energy $E_{\rm form}$ of the vacancy (blue squares) and the work function (chemical potential $-\mu$) of the electrons at the Fermi level (in the vacancy level) for a triple oxygen vacancy cluster at the \chem{TiO_2}-terminated \STO(001) surface. (c) Formation energy diagram for different charge states of the linear oxygen vacancy cluster at the \chem{TiO}-terminated $\STO(001)$ surface as a function of the chemical potential (given with respect to the valence band maximum (VBM).}
		\label{fig:v3b}
	\end{figure}
	
	The charging dependence of significant physical variables similar to that given above for a single vacancy are presented in Fig.~\ref{fig:v3b}. Panel (a) shows the evolution of the valence Bader charge (red circles for average charge $q_{\rm av}$ of the \Ti atoms neighboring the chain [shaded dark blue in Fig.~\ref{fig:dos:v3b}(b)] and pale red rhombs for the charge $q_{\rm c}$ on one of the \Ti atoms neighboring the middle oxygen vacancy) and the cumulative magnetic moment $m$ (blue squares) of the \OV chain as the number of electrons in the calculation supercell is varied. Here we observe a similar tendency as in the case of a single vacancy -- electron depletion (or hole doping) causes a decrease in the localized charge and an ensuing reduction of the magnetic moment. The surplus charge is again predominantly localized at the \Ti atoms surrounding the vacancy chain. Note also how the depletion of vacancy charge affects the edge atoms stronger than it does the central ones [compare the rhombs and the circles in Fig.~\ref{fig:v3b}(a)]. The total magnetic moment is, as said above $m= (6-|n_{\el}|)\mB$. 
	
	Fig.~\ref{fig:v3b}(c) shows the formation energy diagram (similar to that shown in Fig.~\ref{fig:v1}(c) for a single oxygen vacancy) of a triple linear oxygen vacancy cluster at the \chem{TiO}-terminated $\STO(001)$ surface as a function of the chemical potential (given with respect to the valence band maximum (VBM). Here, even clearer than in the case of a single \OV, one can observe a sequence of charge state transitions, starting with a $3\hole \rightarrow 2\hole$ at a chemical potential of $1.25~\eV$ above the valence band maximum and followed by a sequence of transition taking the system through the neutrality to electron doped states. We stress again, that since each state change constitutes a dissipation channel for the external agent responsible for the change, the charge and spin state transitions should be observable experimentally, f.e. within the scope of atomic force microscopy at low temperatures by monitoring the damping of the AFM cantilever oscillations directly related to the dissipation due to the charge and magnetization state changes.
	
	For longer linear vacancy chains the magnetization behavior with charging is expected to be very similar to that presented above. The limiting case of an infinite linear oxygen vacancy chain is shown in the Supplemental Material Fig.~S3 (sketch and PDOS) and Fig.~S4 (formation energy diagram).
	
	At this point we would like to mention a possibility to experimentally sense or measure the presence of charge and magnetic moment transitions in the vacancy quantum dots. To achieve that, we would argue, that, since any change of state or level crossing (with subsequent relaxation of the system) implies a dissipation channel for the external agent provoking the change, the charge and magnetization state changes can be tested experimentally, f.e. within the scope of such dissipation sensitive techniques as atomic force microscopy (AFM) at low temperatures.~\cite{Stomp2005,Cockins2010,Gysin2011,Kisiel2015,Miyahara2017} The relevant quantity hereby would be the damping of the AFM cantilever oscillations directly linked to the magnetic-state-transition induced dissipation in the system.

	Finally we note that motivated by a discussion with experimental colleagues we studied oxygen vacancies at one of the many known reconstruction patterns of a \chem{TiO_2}-terminated \STO(001) surface, namely the $2\!\times\!2$ reconstruction (see Supplemental Material Fig.~S5 and its caption). We find that including the reconstruction into consideration does not alter the main conclusions of the present study -- the lack of an oxygen atom inevitably leads to an excess charge localization on the neighboring Titania and results in a spontaneous magnetization of the latter. Depleting the localization by removing electrons from the surface (f.e. by locally altering the chemical potential) results in a reduction of the magnetic moment of the Titania with its subsequent complete quenching.
	

\section*{Conclusions}

	Oxygen vacancies at (001) surfaces of \STO can be regarded as externally accessible magnetic quantum dots. Their electronic states are determined by \Ti broken bonds that give rise to very localized $d$ states in the gap of the insulating host. More or less like for transition metal impurities, electron-electron interactions give rise to a multiplicity of charge and spin states whose energies are relatively close. The direct exchange coupling of two Ti broken bonds facing each other across the missing \chem{O} surface site is strong and antiferromagnetic, therefore the single vacancy only stabilizes states with low or zero total magnetization. Multiple vacancies offer an even richer scenario, depending on the relative positions of the missing \chem{O} atoms. In this case, an overall state of highest magnetization can also be achieved, as in the linear case which we studied in detail.  Our study of the impurity cluster properties as a function of chemical potential indicates that as in a quantum dot the different surface vacancy states can be tuned and switched  by adjusting the local chemical potential, suggesting their investigation with a local probe tip. The tip-induced transition between different charge and spin states should be traceable in dissipation-sensitive experiments such as the atomic force microscopy.


\section*{Acknowledgements}

	We are very grateful to Marcin Kisiel and collaborators who provided our initial motivation for this research. We gratefully acknowledge the financial support of the ERC Grant No. 320796, MODPHYSFRICT, as well as that of the COST Action MP1303 project, and thank Valeri S. Stepanyuk for the support in terms of calculational resources.
	
\section*{Supplemental Information}
	See Supplemental Material for reference calculations of \STO bulk, list of extended vacancies and surface reconstruction configurations studied as well as the asymptotic example of an extended vacancy - an infinite linear chain - for which a charging manipulation analysis (similar to that presented above for a single and tripple linear vacancy) is given.
	
	
%

	\clearpage
	\onecolumngrid
	\appendix

	\setcounter{figure}{0}
	\renewcommand{\thefigure}{S\arabic{figure}}
	
\section{Bulk $\mathbf{SrTiO_3(001)}$ and single vacancies therein}

	\begin{figure*}[h!]
		\center{\includegraphics{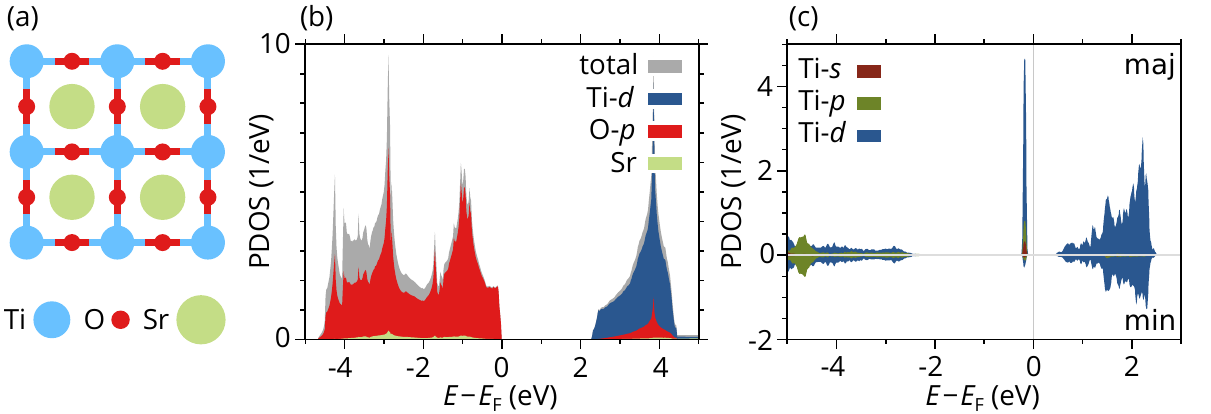}}
		\caption{(a) Sketch of the bulk \STO structure and (b) the relevant partial densities of states making up the electronic structure of the system around the band gap, \textit{i.e.} only the leading orbital contributions are shown: ${\chem{O}}-2p$ (red), ${\Ti}-3d$ (blue) ${\chem{Sr}}-total$. The partial spin-resolved density of states of a Ti atom neighboring a bulk oxygen vacancy in \STO is shown in (c). Positive stands for majority (spin-up) and negative for minority (spin-down) PDOS.}
		\label{fig:dos:bulk}
	\end{figure*}
	
	Bulk \STO [Fig.~\ref{fig:dos:bulk}(a)] calculations were consistent with the state-of-the-art literature. The equilibrium lattice constant is found to be $a_{\rm lat} = 3.99~\AA$, the gap is $2.25~\eV$. It is reduced with respect to the experimental value of $3.25~\eV$ which however does not impact the qualitative message of the present work. The partial atom and orbital projected DOS of the system is shown in Fig.~\ref{fig:dos:bulk}(b).
	
	A single oxygen vacancy calculated including full relaxation of the surrounding atoms yields a ground state with two electrons trapped in the $d$-orbitals of Titania atoms neighboring the vacancy site. The excess charge leads to a spontaneous magnetization of the Titania leaving them with a magnetic moment of $\sim 0.5~\mB$ each, yet the net magnetic moment of the vacancy site is zero, since according to our calculations the two \Ti spins are coupled antiferromagnetically with an exchange energy of about $180~\meV$. The PDOS of the system is shown in Fig.~\ref{fig:dos:bulk}(c).

\section{Extended vacancies at $\mathbf{SrTiO_3(001)}$}

	\begin{figure*}[h!]
		\center{\includegraphics{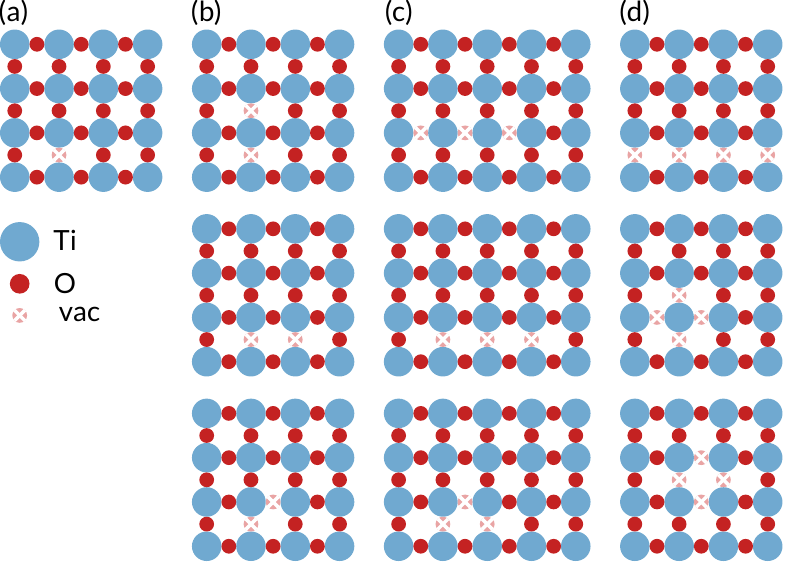}}
		\caption{Studied extended oxygen vacancy clusters at the \STO(001) \chem{TiO_2}-terminated surface -- single (a), double (b), triple (c) and quadruple (d) configurations. Blue and red circles represent the \Ti and O atoms respectively. Sr atoms are not shown in the sketch. Pale red crossed circles mark the vacancy sites. Due to periodicity of the supercell, the topmost configuration in panel (d) actually represents an infinite 1D chain of O vacancies.}
		\label{fig:cfg:ext}
	\end{figure*}
	
	To sample a larger set of potential experimental realizations of surface vacancies we have studied a set of typical configurations of single, double, triple and quadruple 2D oxygen vacancies residing in the topmost layer of the \STO(001) surface. The studied geometries are shown in Fig.~\ref{fig:cfg:ext}. All the studied vacancy clusters exhibit ground state magnetization of \Ti atoms neighboring to \OV sites. Moreover, for most vacancy cluster configurations the coupling between the \Ti atom spins is antiferromagnetic, resulting in either a zero net magnetic moment of the vacancy cluster as a whole or, where the symmetry does not allow for full compensation of the magnetic moment, small met magnetization values. Similarly to the case of a single vacancy, electron depletion causes the vacancy cluster to loose the charge localized therein and the moment of the \Ti atoms to be gradually quenched.

\begin{figure}[h!]
		\center{\includegraphics{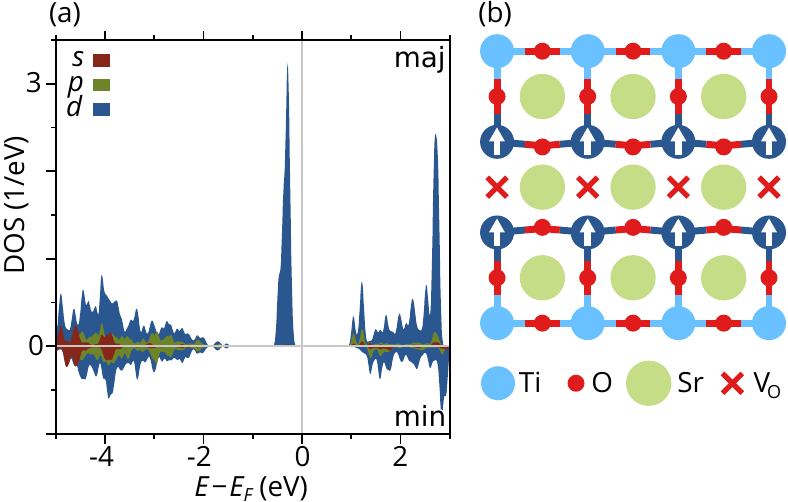}}
		\caption{Partial density of states of one of the \Ti atoms neighboring a linear surface oxygen vacancy chain. Red, green and blue filled curves show the $s$, $p$ and $d$ orbital projected contributions respectively. Positive stands for majority (spin-up) and negative for minority (spin-down) PDOS. The calculations were done with the same unit cell as described in the main text, \textit{i.e.} $4\times 5$ in-plane \STO units.}
		\label{fig:dos:vlin}
	\end{figure}

	A limit case of a infinite linear chain of impurities depicted in Fig.~\ref{fig:cfg:ext}(d, top panel) is in its properties very close the case of a tripple impurity discussed in the text. It exhibits ferromagnetic alignment of the \Ti atom spins with about $2\mB$ per vacancy site in the neutral state, which additively accumulates with increasing chain length. The sketch of the system is presented in Fig.~\ref{fig:dos:vlin}(b) and the partial projected DOS of one of the \Ti atoms neighboring the oxygen vacancy chain is shown in Fig.~\ref{fig:dos:vlin}(a). The vacancy gap state is slightly broadened also here, similar to the case of a tripple \OV discussed in the main text.
	
\begin{figure}[h!]
		\center{\includegraphics{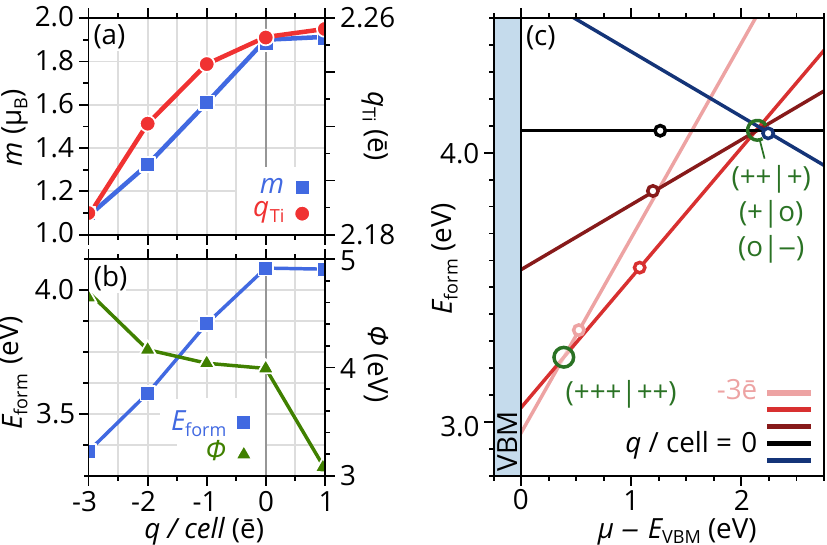}}
		\caption{Electron and hole doping dependence of the (a) valence Bader charge $q_{\Ti}$ (red circles) and cumulative magnetic moments $m_{\Ti}$ (blue squares) of the \Ti atoms neighboring the oxygen vacancy chain, (b) the formation energy $E_{\rm form}$ of the vacancy (blue squares) and the work function (chemical potential $-\mu$) of the electrons at the Fermi level (in the vacancy level) for a quadruple oxygen vacancy cluster at the \chem{TiO_2}-terminated \STO(001) surface. (c) Formation energy diagram for different charge states of a linear oxygen vacancy at the \chem{TiO}-terminated $\STO(001)$ surface as a function of the chemical potential (given with respect to the valence band maximum (VBM). The formation energy is given per unit length of the chain (one \OV vacancy).}
		\label{fig:vlin}
	\end{figure}

	Panel (a) of Fig.~\ref{fig:vlin} shows the evolution of the valence Bader charge (red circles) and the cumulative magnetic moment (blue squares) of the two \Ti atoms surrounding a unit (monomer) of the vacancy chain [note the difference to Fig.~3(a) where the magnetic moment of only one \Ti was presented] as the number of electrons in the calculation supercell is varied. Here we observe a similar tendency as in the case of a single vacancy -- electron depletion (or hole doping) causes a decrease in the localized charge and an ensuing reduction of the magnetic moment.
	
	Fig.~\ref{fig:vlin}(c) shows the formation energy (per oxygen unit) diagram (similar to that shown in Fig.~\ref{fig:v1}(c) for a single oxygen vacancy) of a linear oxygen vacancy chain at the \chem{TiO}-terminated $\STO(001)$ surface as a function of the chemical potential (given with respect to the valence band maximum (VBM). A sequence of charge state transitions can be observed starting with a $3\hole \rightarrow 2\hole$ at a chemical potential of $0.75~\eV$ above the valence band maximum and followed by a rapid sequence of transition around $2.2~\eV$ taking the system through the neutrality to electron doping states. Correcting for the reduced size of the gap in our calculations would likely space out the charge-state transitions making them more distinguishable and better defined.
	
\section{Vacancies at the $\mathbf{2 \times 2}$ reconstructed (001) surface of $\mathbf{SrTiO_3}$}

	\begin{figure*}[h!]
		\center{\includegraphics{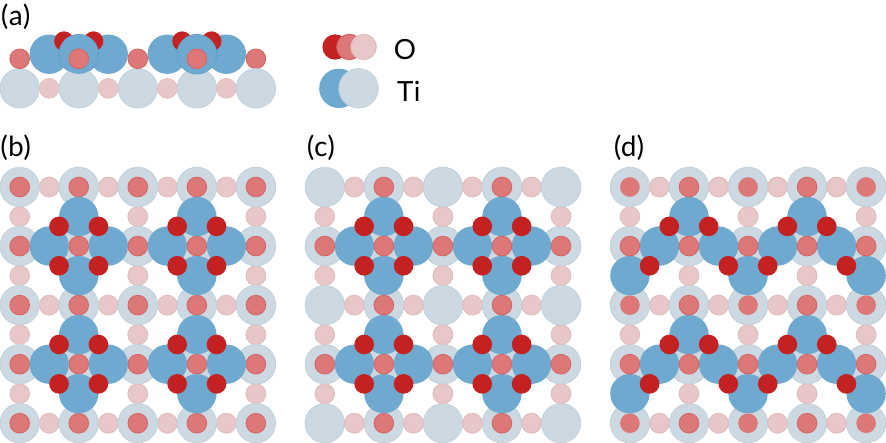}}
		\caption{Studied configurations of a $2\!\times\!2$ reconstructed \STO surface.}
		\label{fig:cfg:2x2}
	\end{figure*}
	
	Motivated by a discussion with experimental colleagues we seek to further increase the applicability of the conclusions attained here by studying the behavior of single oxygen vacancies at a reconstructed \STO(001) surface, choosing as a test subject the example of tree known $2\!\times\!2$ reconstruction patterns. \cite{Shiraki2010,Lin2011a} We find that including the reconstruction into consideration does not alter the conclusions of the present study -- the lack of an oxygen atom inevitably leads to a excess charge localization on the neighboring Titania and results in a spontaneous magnetization of the latter. Depleting the localization by removing electron from the surface (f.e. by locally altering the chemical potential) results in a reduction of the magnetic moment of the Titania with its subsequent complete quenching.
\end{document}